# Demand side management impacts on electricity network vulnerability

Conrad Zorn,[a,b]* Elco Koks,[a,c] Sven Eggimann[a,d]

a   Environmental Change Institute, University of Oxford, South Parks Road, Oxford OX1 3QY, United Kingdom.
b   Department of Civil and Environmental Engineering, University of Auckland, Auckland 1142, New Zealand.
c   Institute for Environmental Studies, Vrije Universiteit Amsterdam, Amsterdam, Netherlands.
d   Urban Energy Systems Laboratory, Swiss Federal Laboratories for Materials Science and Technology, Empa, Dübendorf, Switzerland.

*Corresponding author:* Environmental Change Institute, University of Oxford, South Parks Road, Oxford OX1 3QY, United Kingdom. *E-Mail:* conrad.zorn@auckland.ac.nz

The demand for electricity is undergoing considerable spatial and temporal change. With the uptake of efficient technologies and increased electrification, a better understanding of how potential changes in demand patterns can affect network reliability is necessary. We quantify the macro-economic impacts of potential future changes in demand profiles for an electricity network undergoing generation shortages. Applied to Great Britain, potential savings or losses are assessed for changes to peak demands under four different load profiles: (i) current day, (ii) widespread uptake of efficient appliances, (iii) deployment of heat pumps, and (iv) moving to an idealised fully balanced load profile. Considerable variation in economic disruption is observed both between different demand profiles and across Great Britain. When the networks generation capacity is severely disrupted, we estimate hourly macro-economic impacts to increase by up to £1.23 million per additional GW of national electricity demand. A similar reduction in impacts can be achieved if peak demands are reduced by demand side management. We conclude that risk-related economic impacts are directly linked to the temporal pattern of energy demands and need to be included in the decision-making around demand side measures. We find that decarbonisation strategies without accompanying demand side management may lead to increased economic losses without supply side interventions.

## 1. Introduction

Meeting emissions obligations under the Paris Agreement (UNFCCC 2015) requires significant decarbonisation of the global energy sector (Fell 2017) through a range of socio-technical transformations (Geels et al 2017). While previous attention has largely been focused on supply-side interventions, such as the widespread deployment of low-emission renewable electricity generators and the electrification of carbon based transportation, a better understanding of demand side actions is largely underrepresented despite its recognised importance in literature (Creutzig *et al* 2018, Mundaca *et al* 2018, IPCC 2018, Chaudry *et al* 2015).

For Great Britain, supply-side measures are being targeted in the reconfiguration of a historically fossil-fuel dependent energy sector through increasing renewable electricity generation (Veldhuis *et al* 2018, Barton *et al* 2018, BEIS 2018a). Although about 30% of energy demand is met by renewable generation (BEIS 2018a), it still contributes ~23% of the UK's annual greenhouse gas emissions (BEIS 2017a). Due to the diffusion and uptake of new technologies such as storage technologies, smart grids, and the ongoing electrification of heating and transportation fleets (Qadrdan *et al* 2015), current electricity demand profiles are undergoing considerable temporal and spatial change (Love *et al* 2017, Teng *et al* 2016, Bobmann and Staffell 2015, Grubler *et al* 2018). As a result, Great Britain increasingly relies on local electricity networks with projections of 1 GW additional annual demand from 2030 (National Grid 2017b). Simply responding to emissions targets (Parliament of the United Kingdom 2008) through supply-side generation in isolation is insufficient (Creutzig *et al* 2018, IPCC 2018, BEIS 2017b). In contrast, detailed demand side studies are lacking, particularly those which focus on demand side management at a high spatio-temporal scale (BEIS 2017b, Eggimann *et al* 2019). With demand side management, demands can be shifted away from peak times or reduce overall energy demand and thus change the shape of load profiles (Grunewald and Diakonova 2018, Freeman 2005). However, regardless of the demand profile, functional generation sources and transmission/distribution network assets are critical to meet demands. While Great Britain's electricity transmission network is highly reliable (Espinoza *et al* 2016, OFGEM 2018), a number of recent events have highlighted the need for robust connections to generators. In 2008, unexpected simultaneous shutdowns of two of Great Britain's largest generators led to the tripping of additional generation plants due to drops in grid frequency, ultimately disrupting hundreds of thousands of customers (BBC 2008). In 2015, sudden drops in generation capacity resulted in transmission network operators paying over 40 times the market rate for electricity (Stacey and Adams 2015) and similarly expensive back up capacity was required in 2016 after a boat anchor damaged the France-UK interconnector (Ward 2016). Such events, especially when occurring concurrently, potentially lead to widespread blackouts resulting in significant economic losses particularly in winter periods where supply capacity margins can be as low as 5% (Stacey and Adams 2015).

With many UK-centric studies routinely focusing on quantifying electricity network resilience and reliability following supply-side interventions (Veldhuis et al 2018) or disruptions through natural hazards (Booth *et al* 2017, Espinoza *et al* 2016), this paper seeks to couple potential demand side technological changes with an electricity network model to quantify the impacts of meeting future demand profiles on electricity supply under stressed conditions. To further quantify these impacts we demonstrate the linking of an additional macro-economic model such that the disruptive implications to the wider economy can be compared across different demand scenarios.

Applied to Great Britain, this paper is organised with Section 2 describing the adopted electricity network and the spatio-temporal electricity demand simulation under different demand scenarios, followed by a description of the network disruption approach and coupling with a multiregional economic impact model. Section 3 presents the results of our analysis and compares demand scenarios quantitatively and spatially. We conclude by relating our findings to relevant policy implications and decarbonisation strategies.

## 2. Method

### 2.1 Building electricity network infrastructure

The Great Britain electricity network comprises three voltage groupings connected by transformers: high voltage transmission (400 kV, 275 kV, 132 kV), medium voltage distribution (33kV, 11 kV), and low voltage distribution (230V). To model the Great Britain component of the electricity network (England, Wales, and Scotland) we collate a database of high voltage transmission and medium voltage distribution assets where available, and present these as a graph of nodes with connecting edges (Fig 1a). Nodes represent generation sources, transformer/substation assets, relevant switching points, and demand off-takes. Edges represent overhead lines and cables, both buried and undersea.

The underlying data is assembled from a range of open sources, with a focus on spatial accuracy, rated capacities, generation types, and suitable connection voltages (Bukhsh and McKinnon 2013, SSEN 2018, National Grid 2018, BEIS 2018a). Our collated database of electricity generators presented in Fig. 1a, comprises 2,565 power generation nodes across renewable (37.2 GW), non-renewable (62.2 GW), and international connections (3.75 GW). This equates to over 103 GW of installed rated capacity, estimated to be 82.5 GW when considering realistic output capacity factors (BEIS 2018b, 2018a).

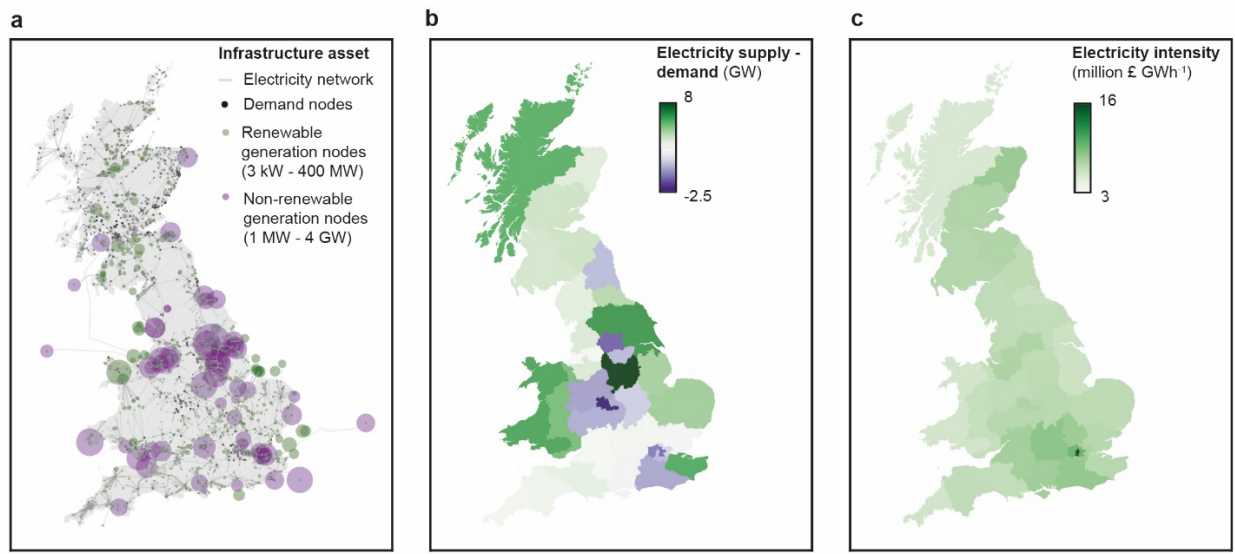

**Fig. 1 | Case study context for Great Britain. a,** Geospatial electricity network representation showing the generation and demand nodes including interconnectors where generation capacities are relative to symbol size. **b**, Capacity margin between local installed generation capacity and peak annual demands across local authority districts. **c,** Spatial distribution of economic activity across NUTS 3 regions in relation to electricity demand.

Major generators are generally located in central and southern areas of GB, with northern (Scotland) generation typically comprising lower capacity renewable generation sources. Relative to demand, the northern areas have excess supply capacity (Fig 1b) which commonly results in a *North-South* flow of electricity towards urban centres. Similarly in Fig. 1c, the electricity intensity, a common measure of productivity per unit energy, sees significant spatial variability with the smallest studied region *Inner London – West* clearly showing the greatest impact on national GDP (European Commission 2018) per unfulfilled GWh of electricity demand. Conversely, the largest region by area, *Scottish Highlands and Islands*, results in the lowest impact per unfulfilled GWh.

Supply-generation edge connectivity across the network given in Fig 1a largely corresponds to the wider national grid, with generated electricity fed into the nearest suitable supply voltage. The feasibility of different loading scenarios are checked using a DC power flow algorithm (Thurner et al 2017). A DC power flow model for active power flow analysis is generally appropriate for high voltage grid analysis (Purchala et al 2005) and as such is similarly assumed in a number of UK-centric studies (Veldhuis et al 2018, Qadrdan et al 2017). Demand nodes are located at the low voltage side of 716 substations ranging from 11 – 400 kV. Properties of line and transformer assets are adopted where known or assumed from standard libraries (Heuck et al 2013, National Grid 2017a, Thurner et al 2017). Further adjustments are made to line and transformer capacities assets such that the network balances under calibrated current day demand profiles as discussed in the following section.

## 2.2 Modelling spatio-temporal electricity demands

To investigate how changes in demand may influence the current energy demand load profile, a set of electricity demand load profiles is simulated. They exemplify a range of potential shifts in load profile patterns due to socio-technical transformation processes. A high resolution energy demand simulation model is used to simulated hourly demands for 379 local authority districts (NUTS 3) for the modelling base year 2015 (Eggimann et al 2019). Hourly electricity data is aggregated over all end uses within each district. Energy demand data for generating the specific load profiles for Great Britain are based on end use and sector specific energy consumption statistics from BEIS (2015). A library of different hourly load profile data is used to disaggregate annual hourly demand.

Four spatio-temporal demand load profiles are simulated based on different assumptions about technological efficiencies and the technology mix for space and water heating, whilst holding all other model parameters constant. For (i) the *heat pump* load profile, 20% heat pumps penetration for space and water heating is assumed. This corresponds to future projected heat pump penetration scenarios up to 2030 (Kreuder and Spataru 2015). For (ii) the *efficiency* load profile, efficiency improvements of appliances across different end uses are assumed. We assume the full realization of efficiency improvements for technologies for space and water heating, lighting, cold, cooking, wet, cooling and humidification and high temperature processes as outlined in Eggimann et al (2019). For (iii) the *heat pump and efficiency* scenario, both heat pump diffusion and efficiency assumptions are combined. In addition to the *current* load profile of the year 2015, we assume a *(vi) flat* load profile with average electricity demand, representing a load profile with no demand variation. Although a flat profile represents an unrealistic situation, it is included here to analyse the idealised case of demand management.

The generated demand profiles from the energy demand simulations are collated to the system level in Fig. 2 for peak demands across the highest demand days (winter, Fig 2a) and lowest demand days (summer, Fig 2b) to represent the possible ranges of network loading.

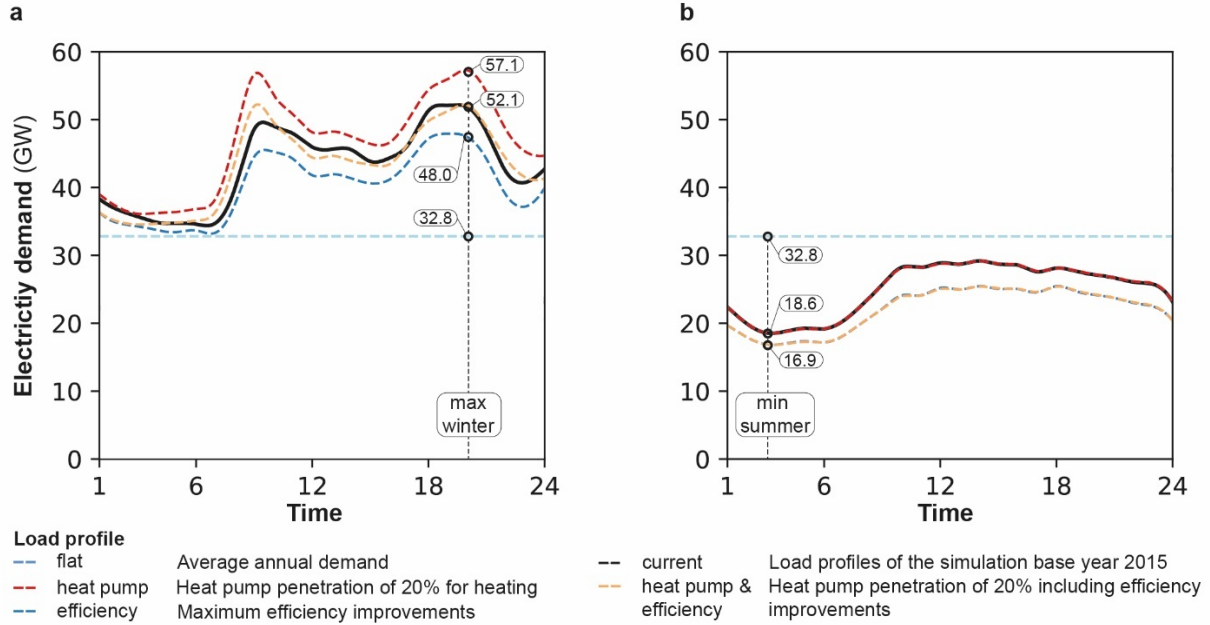

**Fig. 2 | Simulated set of electricity demand load profiles.** Load profiles are shown of the different load profiles for the day with maximum and minimum national electricity demand of the UK. **a**, Days with peak electricity demand occur in winter **b,** days with lowest electricity demands occur in summer.

Under each of the scenarios, the simulated maximum demands (winter period) occur at ~7 pm and minimums (summer period) at ~3 am. The *heat pump* scenario has the largest peak national demand predicted at 57.7 GW compared to the current day at 52.1 GW. As expected, *efficiency* improvements show a reduction in peak demands. The combinations of *heat pumps and efficiency* improvements show similarities with the current day profile in the winter maximum day, and a reduction in demand across the summer period. The *flat* annual average demand profile shows significantly different demands for both the summer and winter periods given the complete removal of fluctuations in demand.

The generated set of load profiles do not include electricity demands for transportation and we did not include sensitivity considerations for different weather scenarios. Whereas decarbonisation is not limited to electrification by heat pumps and further developments affect changes in demand such as the introduction of electric vehicles or switching to a hydrogen-based economy, the motivation behind simulating this particular set of profiles is to generate variability in electricity demand, namely demand increases and decreases.

## 2.3 Simulating network disruptions

To assess the impacts of network disruptions under different demand scenarios, we simulate a loss in generative capacity across the modelled electricity network under each of the demand profiles presented in Fig 2. Firstly, we assemble 1,000 randomly ordered lists of the 2,561 local generation nodes from Fig 1a (excluding international supply points). Secondly, we remove generation nodes to represent a reduction in generative capacity and rebalance the network using an optimised DC power-flow algorithm by minimizing the supply-demand path distances and reliance on international connections (Thurner et al 2017). We assume no solar generation is available regardless of being dropped in each of the scenarios given the peak and minimum demands presented in Fig 2 occurring outside of daylight hours. When demands cannot be fulfilled due to overloaded lines or transformers, demands local to the disrupted generator are dropped from the system until a suitable network state is reached. When such electricity demands are not fulfilled, one can inherently expect economic losses across disrupted sectors of the economy where electricity supply redundancies are not in place. Any reductions in demand are therefore applied as a disruption to the local economy through a multiregional supply-use model for Great Britain as described in the following section.

## 2.4 Economic impact assessment modelling

In order to assess the economic impacts of modelled electricity disruptions, we make use of a multiregional supply-use model, further referred to as the MultiRegional Regional Impact (MRIA) model. For a complete description of the used model, we refer to Koks and Thissen (2016).

The MRIA model allows for estimating a new economic equilibrium as a result of lost economic activity due to power outages. The MRIA model calculates how economic transactions between economic actors may change because of the power outage. Positive and negative economic transactions are considered both within a region and from and to other regions. These transactions (or trade flows) are the main driver of the economic impacts in the affected and surrounding regions. Negative economic impacts will occur when the reduction in production capacity cannot be substituted by other economic actors. Positive impacts may occur if the affected economic actors can find a substitute for either their supply or demand within their existing trade relations. We use multiregional supply-use tables for Great Britain for the year 2013, containing the 36 NUTS 2 regions which are a subset of the regionalised WIOD database of Europe (Thissen et al 2017).

In line with standard input-output modelling, the MRIA model assumes a demand-determined economy, i.e. the total demand must be satisfied by the total supply. The objective function of the MRIA model minimizes total production over all regions. Economic actors aim to minimize their costs given the demand for products and the available production technologies used for their production. These technologies, which can be interpreted as the Leontief production function (Leontief 1936), describe how industries can make a variety of products from a specific set of inputs. The MRIA model is based on the region-specific technologies of industries used to make different products derived from regional technical coefficient matrices. Hence, the technologies are inputs required to produce an output of different products. Production in all regions will take place at the lowest possible costs (industries minimize costs) given intermediate and final demand, the available technologies and the maximum capacity of industries. Finally, the total economic impacts can be interpreted as the change in value added for each industry in each of the 36 regions within GB.

## 3. Results and discussion

Our analysis reveals notable differences across the different load profiles for the simulated electricity network failures with increasing generation losses (Fig. 3a). In the current day scenario, the model shows that ~10% of the available network's generation can be lost before we see any economic impacts. Under the *heat pumps* scenario we observe already the first economic impacts after approximately 3% of removed generative capacity. On the other hand, the *efficiency* scenario only starts to show economic impacts after a ~16% drop in the generative capacity. In each case, this corresponds to impacts being initiated after the residual margin of spare generation capacity is less than 30% of total supply across the combined grid level and embedded generation sources. For the *flat* load profile and summer minimum demands of Fig. 2b, significantly higher losses in generative capacity are apparent in Fig. 3a before demand is unfulfilled and economic impacts arise. In all scenarios, our network model predicts major disruptions and an unstable network after 45% of generative capacity is removed. For this reason, we only consider lost load scenarios of up to 40% lost load, representing a major failing event.

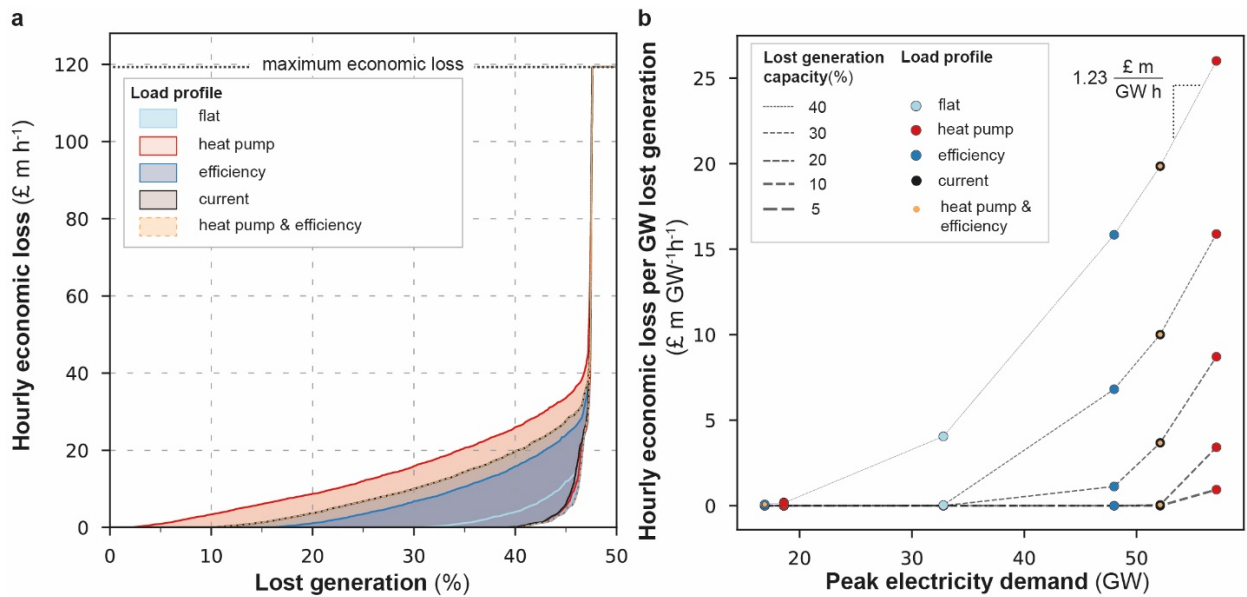

**Fig. 3 | Disruption in relation to electricity load losses and electricity peak demand for Great Britain. a,** The range between minimum and maximum electricity demand for each load profile of the full simulation year are based on 1,000 randomly generated failure combinations and on median cost calculations. The difference in the amount of generation which can be dropped before disruption damage occurs gives an indication of the resilience of each load profile. **b,** The median hourly costs per GW lost load are shown for different lost load scenarios in relation to peak electricity demand. The marginal economic damage is calculated by linear interpolation.

The simulated variability in costs between the minimum and maximum annual demand shows that the economic impacts per unit of lost load varies considerably depending on the hour of the day and the day in the year when network disruptions occurs. Understandably, summer days with low and more constant demands are more resilient to reductions in generative capacity than the winter days given the residual capacities of line and transformer assets across the network.

The marginal costs of disruption are presented in Fig 3b showing the economic impacts per unit increase in demand. These impacts are not consistent across the different scenarios or reductions in generative capacity. As peak demands increase or more generation is dropped from the network, we see greater potential for economic impacts relative to current day demands – an hourly cost of up to £1.23 million per GW of lost load. In contrast, when aggregated demands drop below 18.6 GW, minimal economic disruptions are predicted, even with up to 40% of lost local generative capacity. The implications of this are that future demand profiles should focus on minimising increases to peak demands in preference to avoiding (or reducing) increases to the summertime minimum demands.

Building on these observations, we further investigate the implications of changes to peak demand profiles spatially. Fig. 4 compares the median relative changes in economic impacts for a 40% loss in generative capacity across the modelled network when comparing peak demands to the current peak day scenarios. We focus on the winter peak demand profiles (Fig 2a) to study the network under stressed conditions as we simulated more pronounced differences amongst the simulated load profiles.

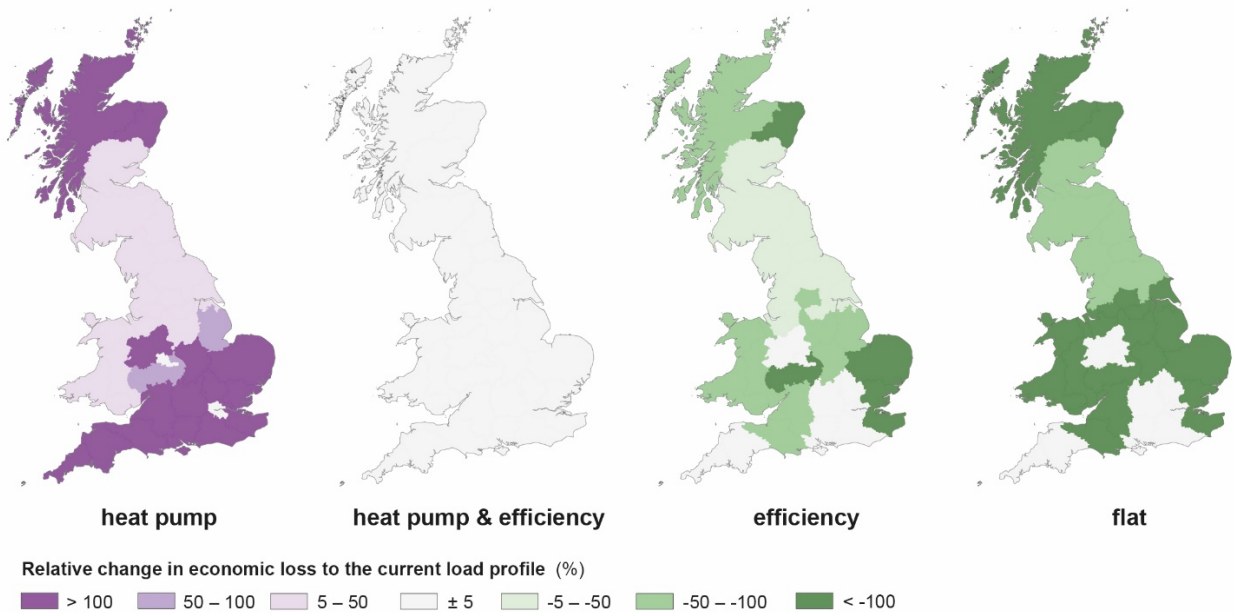

**Figure 4 | Spatial distribution of the relative change in economic impacts for 40% lost generation compared to the current load profile.** Comparing regional median costs to the current load profile shows increasing costs for the heat pump load profile and decreasing costs for the efficiency and flat load profile.

Under the *heat pump* scenario, local authority districts (NUTS3) of the UK see either little change or are worse off when compared to the current day with greater economic impacts observed. This is a result from lower residual capacity left in the network due to increased demands. In comparison, *efficiency* improvements and the *flat* demand profile show widespread reductions in economic impacts for lost reduction capacity given the reduced peak electricity demands. In terms of affected population, in case of the *heat pump* scenario we simulate an increase in economic disruptive costs for 85% of the population when compared to the current day. We see no change for the combined *heat pump and efficiency* load profile. In case of moving towards a fully balanced *flat* profile or improved *efficiencies*, ~62% of the population profit from reduced costs.

## 3. Conclusion

We have demonstrated an explorative modelling approach which can be used to assess economic impact related to network distribution of energy demand management measures on a high spatio-temporal scale. Whereas it is well established that increases in peak electricity demand lead to greater network vulnerability and economic damage resulting from network disruption, its spatio-temporal quantification has been lacking.

A key finding is that with lower peak electricity demand, obtained by various forms of demand side management measures, the economic damages of network disruptions can be considerably reduced. Correspondingly, the potential impact of network distribution on the economy increases when no adequate supply side measures are taken in case of increasing peak electricity demand. We find that per increasing GW, the hourly marginal losses can reach up to £ 1.23 million. This means that working towards more flatter load profiles may provide significant benefits in terms of risk related economic impacts. Our simulation also provides insights into the spatial pattern of losses. We show that not all regions are affected to the same degree, i.e. we simulate for several regions more than twice as high costs between the most and least affected region. Overall, we observe 85% of the population has an increased likelihood of disruption under the same disruption scenario. We conclude that decarbonisation strategies without demand side management could easily lead to increasing economic losses without adequate supply side interventions.

Our findings support that policies to support system flexibility can be considered low regrets options (AURORA 2018), as put forward in recent policy advice (NIC 2018). We argue that increasing peak demand by electrification increases the economic impact of disruptions and needs to be included in the discussion around future heating and transport. This is particularly the case in times of high demand such as winter months where the system is more vulnerable to extreme weather events whilst operating at typically low residual capacity margins.

We provided a strong argument for alternatively exploring demand side management solutions and suggest that actively changing energy demand profiles should be considered as an accompanying strategy to increase the resilience of the energy system in combination with asset robustness and reliability. These findings go far beyond the current discussion which is focused on the provision of additional generation capacity and the building of extra capacity or extra redundancy to electricity networks.

**Acknowledgement.**

This work was supported by the UK Engineering and Physical Science Research Council under grant EP/N017064/1: MISTRAL: Multi-scale InfraSTRucture systems AnaLytics.

**Author contributions**

S.E., C.Z. and E.K. conceived the work and designed the study. S.E. generated energy demand profiles, C.Z. performed the network failure modelling and E.K. performed the economic cost analysis. All authors contributed critically to the draft and gave final approval for publication.